\documentclass[useAMS,usenatbib]{mn2e}
\usepackage{graphicx}
\usepackage{txfonts}
\usepackage{amssymb} 
\usepackage{epsfig} 

\hfuzz=\maxdimen
\def\cha{{\it Chandra}}

\newcommand{\FeTF}{\mbox{{Fe$^{24+}$}}}

\newcommand{\ArXVII}{\mbox{{Ar\,{\sevensize XVII}}}}

\newcommand{\FeXXV}{\mbox{{Fe\,{\sevensize XXV}}}}

\begin{document}

\title[]{Fe K lines in the nuclear region of M82} 

\author[J. Liu et al.]{Jiren Liu$^{1,}$\thanks{E-mail: jirenliu@nao.cas.cn},
	Lijun Gou$^{1}$, Weimin Yuan$^{1}$ and  Shude Mao$^{1,2}$ \\
$^{1}$National Astronomical Observatories, 20A Datun Road, Beijing 100012, China\\
$^{2}$Jodrell Bank Centre for Astrophysics, University of Manchester,
     Manchester, M13 9PL, UK \\
}
\date{}

\maketitle

\begin{abstract}

  We study the spatial distribution of the Fe 6.4 and 6.7 keV lines in
  the nuclear region of M82 using the \cha\ archival data with a total
  exposure time of 500 ks. The deep exposure provides a significant
  detection of the Fe 6.4 keV line.  Both the Fe
  6.4 and 6.7 keV lines are diffuse emissions with similar spatial
  extent, but their morphology do not exactly follow each other.
  Assuming a thermal collisional-ionization-equilibrium (CIE) model,
  the fitted temperatures are around $5-6$ keV and the Fe abundances
  are about $0.4-0.6$ solar value.  We also analyse the spectrum of a
  point source, which shows a strong Fe 6.7 keV line and is
  likely a supernova remnant or a superbubble.  The fitted Fe
  abundance of the point source is 1.7 solar value.
  It implies that part of the iron may be depleted from the X-ray emitting
  gases. If this is a
  representative case of the Fe enrichment, a mild mass-loading
  of a factor of 3 will make the Fe abundance of the point source in
  agreement with that of the hot gas, which then implies
  that most of the hard X-ray continuum ($2-8$ keV) of M82 has a
  thermal origin.  In addition, the Fe 6.4 keV line is consistent with the
  fluorescence emission irradiated by the hard photons from nuclear point
  sources.

\end{abstract}


\section{Introduction}

Galactic-scale outflows (superwinds), driven by stellar winds from
massive stars and core-collapse supernovae (SN) from active
star-forming galaxies, represent an important feedback process of
galaxy evolution \citep[e.g.][]{LH95,Vei05}.  Their mechanical and thermal
energies regulate further star formation and modify the shape of the
galaxy luminosity function \citep[e.g.][]{Ben03}.  The superwinds will
eject metals and enrich halo gases and intergalactic medium
\citep[e.g.][]{Son97,Pet01,Tum11}.  However, due to its high
temperature, low density, and especially the contamination of 
dense point sources residing in the star-forming region, the hot
plasma driving superwinds is hard to observe.

The prototype starburst galaxy M82 (located at 3.6 Mpc), with a
powerful superwind detected on scales up to 10 kpc, is an ideal target
to study the driving plasma of superwinds.  With the sub-arcsec
spatial resolution of \cha, \citet{Gri00} detected diffuse hard X-ray
emission in the nuclear region of M82. They also detected an emission
line around $5.9-6.9$ keV, which is likely due to the \FeXXV\ K$\alpha$ 
line at 6.7 keV.  This is expected in the scenario of
superwinds \citep{CC85}, which predicts a metal-enriched hot plasma at
temperatures of $10^7-10^8$ K.  The corresponding Fe abundance is
about 0.3 solar abundance if assuming a thermal spectrum.
\citet{Str07} examined and discussed Fe lines of M82 in detail.  They
found that the Fe 6.7 keV line luminosity is consistent with that
expected from the enrichment of previous SN ejecta. They also reported
a marginal detection of the Fe 6.4 keV line, which is a fluorescent
line of the neutral-like Fe.

The Fe 6.7 keV line is important to understand the driving plasma of superwinds. 
It is the most prominent emission line at such high temperatures and thus the 
best tracer of the metal enrichment.  
If all the Fe produced by massive stars are mixed in the hot plasma,
the expected Fe abundance will be around 5 times solar abundance, in contrast to the
observed 0.3 times solar abundance \citep{Str07}.  It then implies there is other
contribution to the diffuse continuum, in addition to the thermal continuum,
or part of the Fe is depleted from the X-ray emitting gases.

In previous studies, the nuclear region of M82 is taken as a whole and no spatial analysis
has been done due to the limited statistics (the total exposure is less than 50 ks).  In
this letter, we present a detailed spatial study of the Fe 6.7 keV line of M82 with 
500 ks of \cha\ archival data, which is a factor of 10 longer than that used in
previous studies.  The deep data allows a significant detection of the Fe 6.4 keV line,
which is informative to study the coexistence of cold molecular and hot gases.  We also
report a point source showing the Fe 6.7 keV line, which provides strong 
implications for the mixing level of Fe produced by massive stars.

The paper is structured as follows. We describe the data reduction in \S 2 and the
analysis results in \S 3.  The implications of the results are discussed in \S 4.  The
errors quoted are for the 90\% confidence level throughout the paper.

\begin{table}
\caption{List of \cha\ observations used for the analysis}
\tabcolsep 3.8pt
\begin{tabular}{lccccl}
\hline
ObsID &  $t_{\rm tot}$ (ks)& $t_{\rm eff}$(ks)& R.A. & Dec. & Obs time\\
\hline
10542 &	118 & 116&09:55:51.3 & +69:42:51.6 & 2009-06-24\\	
10543 &	118 & 110&09:55:37.6 & +69:42:25.1 &2009-07-01\\	
10544 &	74  & 67&09:55:54.2 & +69:38:57.7 &2009-07-07\\	
10925 &	45  & 40&09:55:54.2 & +69:38:57.7 &2009-07-07\\	
10545 &	95  & 90&09:56:07.8 & +69:39:34.1 &2010-07-28\\	
11800 &	17  & 17&09:56:07.8 & +69:39:34.1 &2010-07-20\\	
5644  &	68  & 49&09:55:50.2 & +69:40:47.0 &2005-08-17\\	
2933 &	18  & 16&09:55:52.6 & +69:40:47.1 &2002-06-18\\	
\hline
\end{tabular}
\begin{description}
\begin{footnotesize}
\item
Note: $t_{\rm tot}$ is the total exposure time and $t_{\rm eff}$ is 
the effective exposure time with the period of flares removed.
\end{footnotesize}
\end{description}
\end{table}

\begin{table*}
\caption{Spectral fitting results.}
\label{tab3}
\begin{tabular}{@{}ccccccccccc@{}}
\hline
& \multicolumn{4}{c}{{\sc Vapec+Gauss}} & &\multicolumn{5}{c}{{\sc Power-law+Gauss+Gauss}}  \\ \cline{2-5} \cline{7-10}
  Region   & kT &$Z_{\rm Fe}$ & Norm&6.4 keV line & $\chi^2_{\nu}$ &  $\Gamma$ & Norm  &6.4 keV line&6.7 keV line &$\chi^2_{\nu}$ \\ 
\hline
1 
&5.2$^{+1.2}_{-0.8}$&0.40$^{+0.15}_{-0.14}$&2.9$^{+0.3}_{-0.3}$ &4.2$^{+2.0}_{-2.0}$ &0.85& 2.2$^{+0.2}_{-0.2}$ &1.5$^{+0.4}_{-1.4}$ & 4.0$^{+2.1}_{-2.0}$ & 6.7$^{+2.5}_{-2.5}$  & 0.85\\
2
&6.3$^{+1.1}_{-0.8}$&0.57$^{+0.14}_{-0.13}$&4.6$^{+0.4}_{-0.3}$ &6.4$^{+2.6}_{-2.6}$ &0.75& 2.0$^{+0.2}_{-0.1}$ &2.0$^{+0.5}_{-0.3}$ & 6.0$^{+2.7}_{-2.6}$ & 13.5$^{+3.4}_{-3.5}$ & 0.77\\
3
&6.4$^{+1.3}_{-1.0}$&0.37$^{+0.14}_{-0.13}$&3.4$^{+0.3}_{-0.3}$ &3.4$^{+2.2}_{-2.2}$ &0.82& 2.0$^{+0.2}_{-0.1}$ &1.5$^{+0.4}_{-0.3}$ & 3.1$^{+2.2}_{-2.1}$ & 7.1$^{+2.7}_{-2.7}$  &0.80\\
4
&4.1$^{+0.9}_{-0.6}$&0.52$^{+0.18}_{-0.16}$&2.6$^{+0.3}_{-0.4}$ &0.4$^{+0.4}_{-0.4}$ &1.18& 2.5$^{+0.2}_{-0.2}$ &1.7$^{+0.6}_{-0.4}$ & 0.3$^{+0.7}_{-0.3}$ & 7.2$^{+2.4}_{-2.4}$  & 1.09\\
5
&5.4$^{+0.8}_{-0.5}$&0.53$^{+0.13}_{-0.12}$ &4.4$^{+0.3}_{-0.3}$ &4.8$^{+2.3}_{-2.3}$&1.01 & 2.2$^{+0.2}_{-0.1}$&2.3$^{+0.6}_{-0.4}$  & 4.6$^{+2.4}_{-2.3}$ & 12.7$^{+3.2}_{-3.3}$ & 0.97 \\
6
&10.4$^{+2.9}_{-2.7}$&0.50$^{+0.33}_{-0.29}$ &2.1$^{+0.2}_{-0.1}$& 2.4$^{+1.9}_{-2.0}$&0.93& 1.8$^{+0.2}_{-0.2}$&0.7$^{+0.2}_{-0.2}$  & 2.3$^{+2.0}_{-1.9}$ & 3.0$^{+2.2}_{-2.2}$& 0.97 \\
A                 
&5.8$^{+2.0}_{-1.3}$&1.72$^{+0.65}_{-0.54}$ & 0.9$^{+0.1}_{-0.2}$ & -- &1.08 & --&--&--&--&--\\

\hline 
\end{tabular} 
\begin{description} 
\begin{footnotesize} 

\item Note: kT is in units of keV; the Fe abundance $Z_{\rm Fe}$ is relative to the solar
value; the line intensity is in units of $10^{-7}{\rm photons\ cm^{-2}s^{-1}}$; ${\rm
Norm}\times10^{-4}$ is the normalization for the {\sc Xspec} models,
{\sc Vapec} and {\sc Power-law},
respectively;  $\Gamma$ is the photon index of the power-law model. 

\end{footnotesize}
\end{description} \end{table*}

\section{Observation data}

We use 8 archival \cha\ observations (ObsID 5644 by PI T. Strohmayer and the
others by PI D. Strickland) listed in Table 1, all of which are observed with ACIS-S3. 
After removing the period of flares, the effective exposure time is about 500 ks, which
allows a detailed spatial study of Fe lines.  The data reduction is performed using the
CIAO software (version 4.5).

The nuclear region of M82 is divided into six box regions with the
same size ($12''\times10''$ each) based on the photon count rate between
3.3 and 7.5 keV as illustrated in Figure 1.  The adjacent line between
regions 1, 2, 3 and regions 4, 5, 6 is along the major axis of the
stellar disk of M82.  The excellent angular resolution of \cha\ can
help to resolve bright point sources, which are the key contamination
for studies of the diffuse emission.  The point
sources are detected using the tool {\it wavdetect} in CIAO with an
encircled psf fraction of 90\% and a size parameter {\it ellsigma} of
3. Because the nominal pointing direction is different for different
observations, the source detection has been done separately.  As a
demonstration, Figure 1 shows the excluded point sources marked with
ellipses for ObsID 10542.  The typical radii are around
1.5$''-2''$. For the central two bright sources, their radii are
enlarged by a factor of 2 to minimize their contamination.

For an extended source like M82, it is generally hard to find
emission-free regions to do the background subtraction. We use the blank-sky
datasets produced by the ACIS calibration
team\footnote{http://cxc.harvard.edu/ciao/threads/acisbackground/} to
estimate the background, which is extracted
from the same CCD region for the given box.

\section{Analysis results}
\subsection{Diffuse region}
\begin{figure}
\includegraphics[height=2.20in]{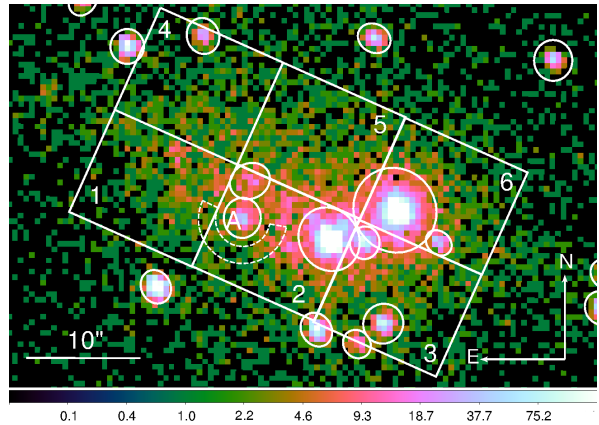}

\caption{Spectral extraction regions over-plotted on the $3.3-7.5$ keV
  counts map of M82 for ObsID 10542.  The ellipses show the excluded
  point sources for ObsID 10542.  The point source indicated by the
  letter 'A' shows the Fe 6.7 keV line and is discussed in \S 3.2. The
  dashed panda is the background region adopted for Source A.  }

\end{figure}

\begin{figure*}
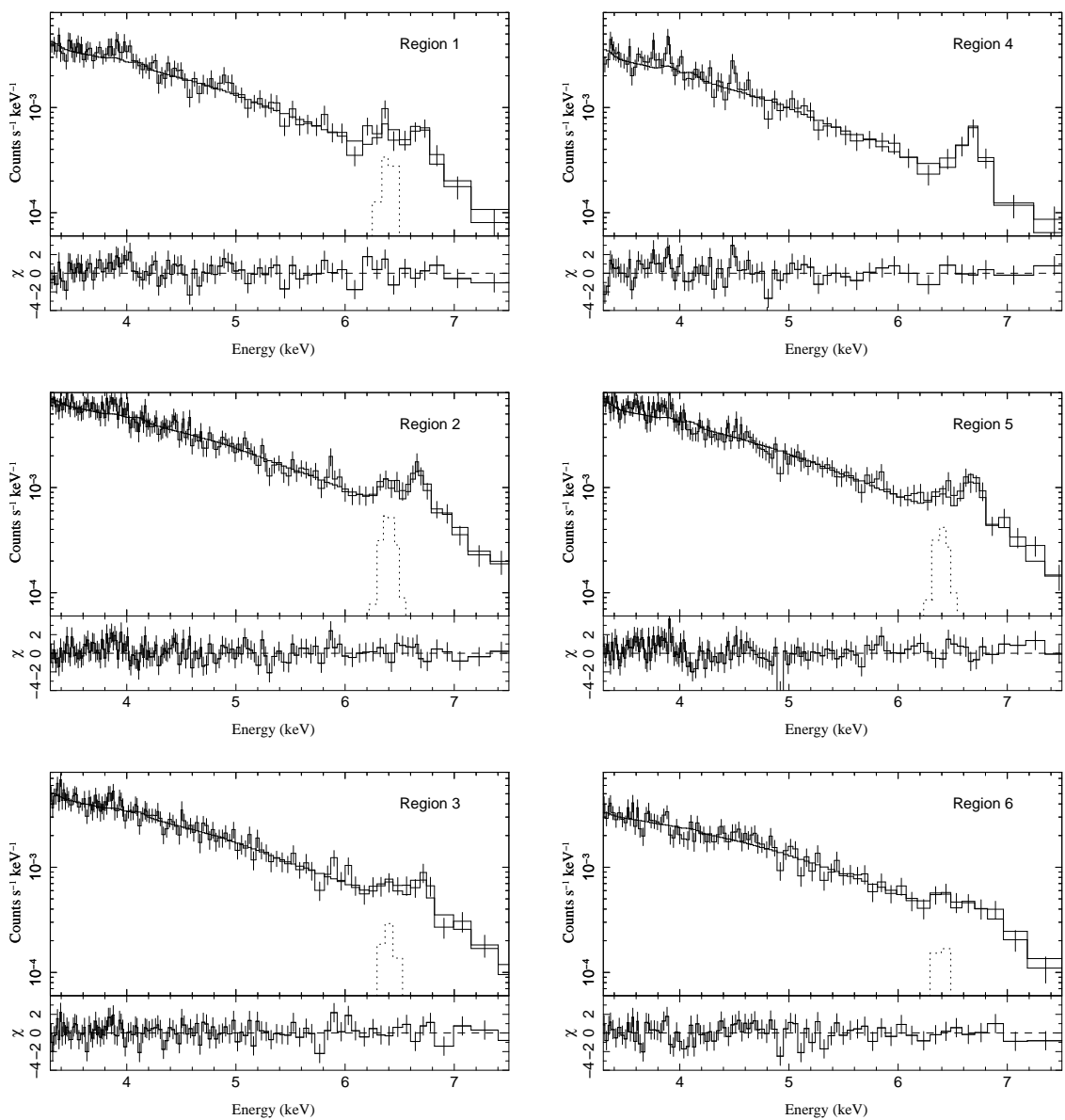

\includegraphics[height=2.10in]{V1.ps}
\includegraphics[height=2.10in]{V4.ps}
\includegraphics[height=2.10in]{V2.ps}
\includegraphics[height=2.10in]{V5.ps}
\includegraphics[height=2.10in]{V3.ps}
\includegraphics[height=2.10in]{V6.ps}
\caption{
Background-subtracted spectra of six box regions combined for all eight 
observations.
The fitting results of the CIE model plus a Gaussian line centered at 
6.4 keV are plotted as solid histograms. The fitted Fe 6.4 keV line is 
indicated as dotted line separately. $\chi$ is the difference between data
and model divided by the error.
} 
\label{SEplot}
\end{figure*}

The background-subtracted spectrum, combined from all eight observations
for each region, is plotted in Figure 2. The Fe 6.4 and 6.7 keV
lines are clearly seen in most regions.  We find the nuclear spectra
of M82 within $0.5-7.5$ keV can be fitted with two thermal
collisional-ionization-equilibrium (CIE) models with temperatures
around 0.65 and $5-6$ keV, respectively.  The contribution of the
low-temperature component is about 5\%
around $3-4$ keV and negligible for higher energies.  Thus, for the
purpose of the study of the Fe 6.4 and 6.7 keV lines, we limit the
fitting energy range to $3.3-7.5$ keV and apply only one thermal CIE model
\citep[{\sc Vapec},][]{Apec12} plus a Gaussian line to fit the spectrum. 
As the Fe lines are the only strong lines in the fitting range, we 
only fit the Fe abundance in the {\sc Vapec} model, and the abundances
of other elements are fixed to solar values \citep{Lod03}.
The Gaussian line is centered at 6.4 keV
and its line width is set to a minimum of 10$^{-6}$ keV,  as it 
can not be reliablely measured due to the limited instrument resolution.
To avoid the contamination of the emission lines of \ArXVII,
the energy range between $3.8-4.0$ keV is also ignored. 
The data are binned to have a minimum count of 25.  
The spectral analysis is done with the ISIS package \citep{ISIS}, which 
calls the Xspec models \citep{Arn96}.

The fitting results are plotted in Figure 2 and listed in
Table 2.  From Figure 2 we see that the CIE model plus a Gaussian line
generally provides a reasonable fit to the observed spectra.
The fitted temperatures are around $5-6$ keV, except for region 6,
which may be contaminated by the brightest point source of M82.
The fitted Fe abundances are around $0.4-0.6$ times
solar value.  
To measure the intensity of Fe 6.7 keV line, we refit the spectra
using a power-law model plus two Gaussian lines centered at 6.4 keV
and 6.7 keV, respectively.  The fitting results are also listed in
Table 2.  The fitted power-law indices are around 2.  The goodness of
the fit for the power-law model is similar to that for the CIE
model. The 6.7 keV line is generally stronger than the 6.4 keV line.

\begin{figure}
	\hspace*{-0.5cm}
\includegraphics[height=2.2in]{objA.ps}
\caption{Background-subtracted spectrum of the point source A (see Figure 1)
that shows the Fe 6.7 keV line.
The fitted thermal CIE model is plotted as the solid histogram.}
\label{NWplot}
\end{figure}

\begin{figure*}
\includegraphics[height=2.5in]{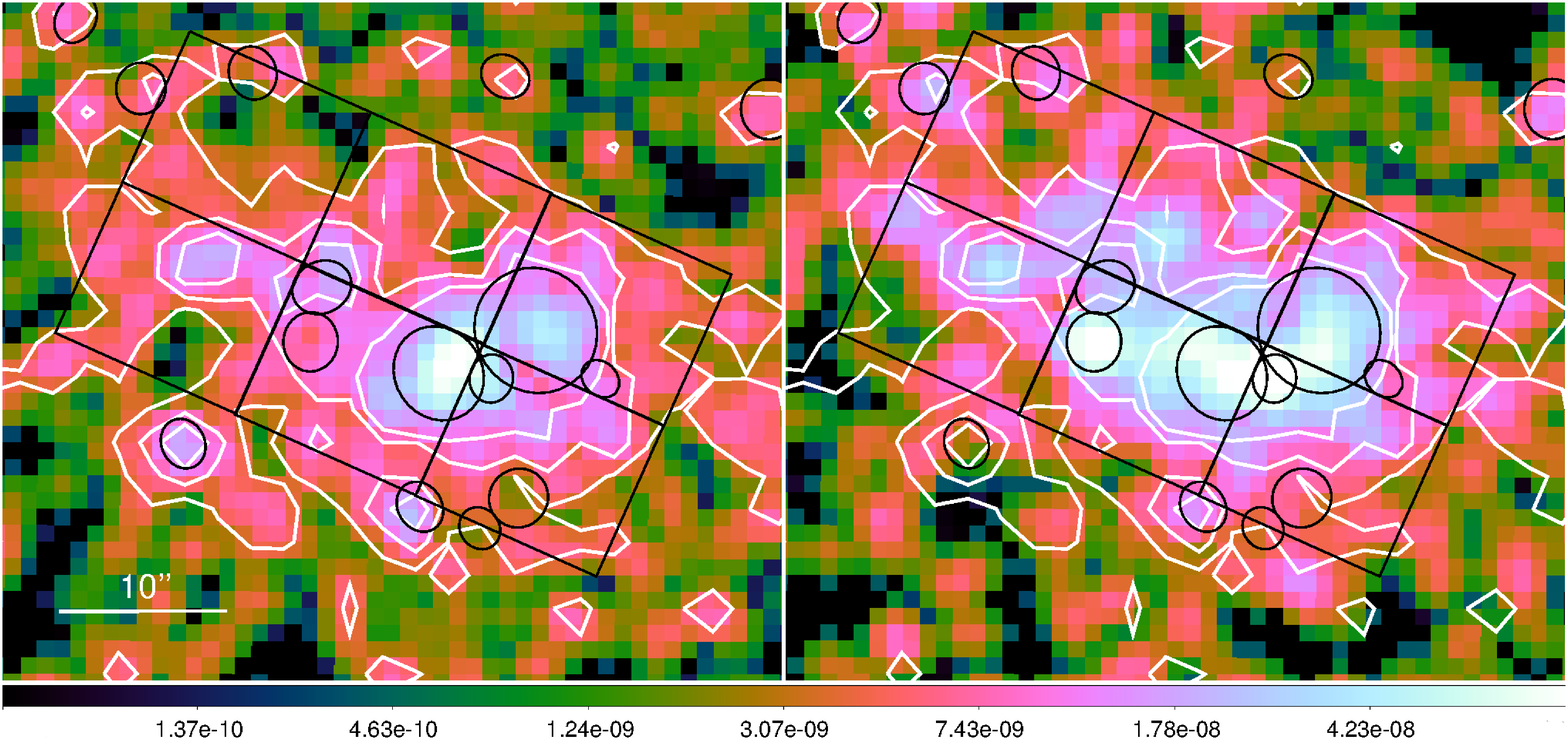}
\caption{Flux maps of photons within $6.25-6.55$ keV (left) and $6.55-6.85$ keV 
(right), combined from all eight exposures. 
A continuum of the linear interpolation of maps of
$5.95-6.25$ keV and $6.85-7.15$ keV at 6.4 and 6.7 keV has been subtracted
respectively. The exposure-map of the central energy of each band is used 
when making the flux map. The maps are binned to 0.984 arcsec and 
smoothed with a two-pixel Gaussian function. 
The color-bar shows the flux in units of photons$^{-1}$cm$^{-2}$s$^{-1}$pixel$^{-1}$.
The white contours of the $6.25-6.55$ keV map are plotted in both panels. 
The extraction regions and excluded sources (black) in Figure 1 are also plotted.
}
\end{figure*}

In the fitting above we have neglected the contribution from the
unresolved X-ray point sources. \citet{Gri03} found that the
differential luminosity function of the X-ray point sources in
star-forming galaxies follows the form of
$N(L_{38})=3.3L_{38}^{-1.6}$SFR, where $L_{38}$ is the $2-10$ keV
X-ray luminosity in units of $10^{38}$ erg s$^{-1}$ and SFR is the
star formation rate measured in units of M$_{\sun}$\ yr$^{-1}$.  The
luminosity of the faintest source in the analysed regions is about
$4\times10^{37}$ erg s$^{-1}$ in $3.3-7.5$ keV, which corresponds to
$10^{38}$ erg s$^{-1}$ in $2-10$ keV for a power-law model with a photon
index of 1.8.  The adopted SFR of M82 in \citet{Gri03} is 3.6 M$_{\sun}$\ yr$^{-1}$ 
based on the H$\alpha$ flux of the entire galaxy.
Assuming that half of the X-ray point sources are located at the analysed regions
(a fraction estimated from the detected point sources), 
for unresolved sources between $5\times10^{35}$ erg s$^{-1}$ (the lower limit of the 
point sources for the luminosity function of \citealt{Gri03}) and 
$4\times10^{37}$ erg s$^{-1}$ (the lowest luminosity of the detected sources), 
the contribution to the diffuse emission is about 20\%. 
We test the effect of the unresolved sources by including a power-law
model with a photon index of 1.8, the contribution of which is fixed at
20\%.  The fitted temperatures is a little lower, but still in
the ranges obtained without considering the unresolved sources. The Fe
abundances are increased by about 20\% compared with the results without
the power-law model.

\subsection{Point sources}

We extract the spectra from all individual point sources detected in
the nuclear region of M82 and find two sources showing the Fe 6.7 keV
line but none showing the Fe 6.4 keV line.  One source only
appears in the 2010 exposures (ObsID 10545 and 11800) and because
its statistics is too low to provide a reliable analysis, it is
discarded here.  The other one is indicated by the letter 'A' in Figure 1.  It is
spatially correlated with the radio source 44.0+59.6 detected by
\citet{Hua94} and is possibly a SNR or a superbubble produced by
several SNRs.

The spectrum of Source A is plotted in Figure 3, for which a
background extracted from the dashed panda region illustrated in
Figure 1 is subtracted.  We also fit the spectrum of source A with a
CIE model.  The fitting result is listed at the bottom row of
Table 2.  The fitted temperature is 5.8 keV, similar to that measured
in the diffuse regions. It corresponds to a shock
velocity about 2000 km/s \citep[e.g.][]{Vink12}.  
The flux of source A between 2 and 8 keV is $1.4\times10^{38}$ erg s$^{-1}$.
The fitted Fe
abundance is 1.7 times solar value, which is about 3 times that of the
diffuse regions. Its implications are discussed in \S 4.

\subsection{Imaging analysis}

To further illustrate the spatial distribution of the Fe 6.4 and 6.7 keV
lines, we plot the flux map of photons within $6.25-6.55$ keV and
$6.55-6.85$ keV in Figure 4. We see that both emissions have diffuse
morphology and similar spatial extent. The $6.55-6.85$ keV map is 
slightly more extended along the minor axis of the disk than the
$6.25-6.55$ keV map.  It is also clear that the two maps do not
exactly follow each other and some regions with high $6.25-6.55$
keV flux show less $6.55-6.85$ keV flux.

\section{Discussion and Conclusion}

We conduct a spatial study of the Fe 6.4 keV and 6.7 keV lines in the
nuclear region of M82.  The Fe 6.4 keV line is clearly detected with
the deep datasets.  The emission of both lines have similar spatial
extent, but their morphology do not exactly follow each other.
The total luminosity of the Fe 6.7 keV line is about $1\times10^{38}$
erg s$^{-1}$, which is consistent with the value measured by
\citet{Str07}.  The total luminosity of the Fe 6.4 keV line is about
$4\times10^{37}$ erg s$^{-1}$.

The spectra can be fitted well with a CIE model plus a Gaussian line 
over the energy range of $3.3-7.5$ keV.
The fitted temperatures are around $4-6$ keV, which 
are consistent with the scenario of superwinds \citep{CC85}.  
The fitted Fe abundances are around 0.5 solar value.  
\citet{Str07} calculated the Fe abundance produced by
SN ejecta and stellar winds using {\it Starburst99} and found 5 times
solar abundances.  However, the iron may not be well mixed with the
hot plasma.  The spectrum of the point source A provides an Fe
abundance of 1.7 times solar value.  Although source A can not be the
population of SNRs that are responsible for the present hot gas, if it
represents a typical case, then the iron may be depleted heavily. A
mild mass-loading of a factor 3 will make the Fe abundance of source A
in agreement with that of the hot gas. 

As discussed by \citet{Str07}, the inverse Compton spectrum will have
a photon index around 1.3 if the electron population is also
responsible for the synchrotron radiation. This index is different
from our fitting results of the power-law models, which have photon
indices around 2.  The contribution of a power-law
model of index of 1.3 can not exceed 50\%, otherwise it will produce more
flux than observed at energies above 7 keV.
Adding a power-law model of index of 1.3 to the
fit will lower the fitted temperatures.  If we require the
fitted temperatures to be above 4 keV, the contribution of the inverse
Compton emission can not exceed 30\%. This is consistent with the
contribution fraction of 25\% estimated by \citet{Str07}.  Including
both the contributions of unresolved point sources and inverse Compton
emission will increase the Fe abundances to $0.7-0.9$ solar value, still far below
the expected value if there is no depletion of Fe.


We have assumed thermal CIE models when fitting the Fe abundances in
\S 3.  The ion at the charge state i approaches to ionization
equilibrium on a timescale of $t \sim [n_e(C_i+\alpha_i)]^{-1}$, where
$n_e$ is the electron density, $C_i$ is the collisional ionization
rate and $\alpha_i$ is the recombination rate out of the charge state
i \citep{L99}. For a temperature of $6\times10^7$ K, $t \sim
1.5\times10^4 n_e^{-1}$ yr cm$^{-3}$ for the ion of \FeTF. 
The electron density can be estimated from the normalization
of {\sc Vapec} model, which is $\frac{10^{-14}}{4\pi D^2}\int n_en_H\,dV$,
where $D$ (in units of cm) is the distance to M82, $n_e$ and $n_H$ are 
the electron and hydrogen densities in units of cm$^{-3}$.
We adopt a depth of 0.6 kpc, which is the disk spatial extent of the 
X-ray emitting region of M82.
Assuming a filling factor of 1 and $n_H=0.8n_e$, the electron density 
$n_e\sim0.35$ cm$^{-3}$ for the hot gas 
in the nuclear region of M82.  This gives
an ionization equilibrium timescale of $4\times10^4$ yr.  It is
relatively short compared with the typical timescale ($\sim10^6$ yr) in the
starburst region and the outflow timescale ($\sim5\times10^5$ yr) of
the nuclear region of M82. It suggests that non-equilibrium-ionization
is unlikely to be important for the Fe 6.7 keV line of M82.  When
applying a non-equilibrium ionization model ({\sc Vnei}), we
find no improvement in the fits, and it provides a little higher
temperature ($7-8$ keV) and a lower Fe abundance.

The Fe 6.4 keV line is the fluorescent line of neutral-like Fe.  It
has also been observed in another starburst galaxy of NGC 253
\citep{Mit11}.  They attributed the 6.4 keV line to the irradiation of
molecular gas by surrounding point sources. A similar mechanism is
applicable for M82, the nuclear region of which contains plenty of
cold molecular gases \citep[e.g.][]{Wei01}.

The luminosity of all nuclear point sources within $7-8$ keV is about
$5\times10^{39}$ erg s$^{-1}$, which is about 15 times the $7-8$ keV
luminosity of the diffuse emission and 100 times the luminosity of the
Fe 6.4 keV line.  The H$_2$ column densities in the nuclear region of
M82 are measured to be around $5\times10^{22}$ cm$^{-2}$
\citep{Wei01}. Assuming a solar Fe abundance ($3\times10^{-5}$
relative to the number of H), the absorption optical depth of the
neutral Fe is about 0.1 at 7 keV given an absorption cross-section of
$4\times10^{-20}$ cm$^2$ \citep{Vei73}.  Taking into account the
fluorescence yield of the neutral Fe K line of 0.34
\citep[e.g.][]{Kal04}, the observed Fe 6.4 keV line luminosity
($4.5\times10^{37}$ erg s$^{-1}$) is consistent with the irradiation
by the hard X-ray photons from nuclear point sources.

\section*{Acknowledgments}
We thank the referee for his/her detailed and valuable report and
Youjun Lu for helpful discussions.  This work is supported by National
Natural Science Foundation of China for Young Scholar (grant
11203032). We also thank the Chinese Academy of Sciences and NAOC for
support.  It is based on observations obtained with the \cha\ X-ray
observatory, which is operated by Smithsonian Astronomical Observatory
on behalf of NASA.

\end{document}